\documentclass[12pt]{article} 

\usepackage{helvet}

\newcommand*{\myEXPfont}{\fontfamily{cmr}\selectfont}

\usepackage{graphicx} 
\usepackage{mathtools}
\usepackage{amsmath}
\usepackage{multirow}
\usepackage[export]{adjustbox}
\usepackage[left=1in, right=1in, top=1in, bottom=1in]{geometry} 
\usepackage{booktabs}
\usepackage{rotating}
\usepackage{rotfloat}
\usepackage{float}
\usepackage{subfig}
\usepackage{url}
\usepackage{pdfpages}
\usepackage{rotating}
\usepackage[doublespacing]{setspace} 
\usepackage{csvsimple}
\usepackage{array}
\usepackage{wrapfig}
\usepackage[utf8]{inputenc}

\usepackage{cite}
\usepackage{tabularx}
\usepackage{float}
\usepackage{verbatim}

\usepackage{amsthm}                
\usepackage{amssymb}

\usepackage{multirow}

\usepackage{caption}

\usepackage{amsfonts}
\usepackage{amsmath} 
\usepackage{tikz}
\usetikzlibrary{positioning,shapes.geometric}

\usepackage{xcolor}

\usepackage{setspace}

\DeclareMathOperator{\E}{\mbox{{\myEXPfont E}}}

\usepackage{mathrsfs}

\makeatletter
\newcommand*{\indep}{%
  \mathbin{%
    \mathpalette{\@indep}{}%
  }%
}
\newcommand*{\nindep}{%
  \mathbin{
    \mathpalette{\@indep}{\not}
  }%
}
\newcommand*{\@indep}[2]{%
  \sbox0{$#1\perp\m@th$}
  \sbox2{$#1=$}
  \sbox4{$#1\vcenter{}$}
  \rlap{\copy0}
  \dimen@=\dimexpr\ht2-\ht4-.2pt\relax
  \kern\dimen@
  {#2}%
  \kern\dimen@
  \copy0 
} 
\makeatother

\definecolor{forestgreen}{RGB}{34,139,34}
\usepackage{hyperref}
\hypersetup{
    colorlinks=true,
    linkcolor=magenta,
    filecolor=magenta, 
    citecolor=forestgreen,      
    urlcolor=blue
}

\usepackage{sectsty}
\sectionfont{\large}

\usepackage{layout}

\usepackage{geometry}

\newcolumntype{C}[1]{>{\centering\arraybackslash}p{#1}}

\usepackage{authblk}

\usepackage{afterpage}
\usepackage{lscape}
\usepackage{graphicx}

\usepackage{datetime}

\usepackage{amsthm}

\raggedright
\begin{document}

\title{\textbf{Generalizability analyses with a partially nested trial design: the Necrotizing Enterocolitis Surgery Trial} \vspace*{0.3in} }

\author[1,2]{Sarah E. Robertson}
\author[3]{Matthew A. Rysavy}
\author[4]{Martin L. Blakely}
\author[5]{Jon A. Steingrimsson}
\author[1,2,6]{Issa J. Dahabreh\thanks{Address for correspondence: Dr. Issa J. Dahabreh, MD ScD; Department of Epidemiology; Harvard T.H. Chan School of Public Health, Boston, MA 02115; email: \href{mailto:idahabreh@hsph.harvard.edu}{idahabreh@hsph.harvard.edu}; phone: +1 (617) 495‑1000.}}

\affil[1]{CAUSALab, Harvard T.H. Chan School of Public Health, Boston, MA}
\affil[2]{Department of Epidemiology, Harvard T.H. Chan School of Public Health, Boston, MA}
\affil[3]{Department of Pediatrics, McGovern Medical School, The University of Texas Health Science Center, Houston, TX}
\affil[4]{Department of Pediatric Surgery, Vanderbilt University Medical Center, Nashville, TN}
\affil[5]{Department of Biostatistics, Brown University School of Public Health, Providence, RI}
\affil[6]{Department of Biostatistics, Harvard T.H. Chan School of Public Health, Boston, MA}


\maketitle{}
\thispagestyle{empty} 

\newpage 


\vspace{0.1in}
\noindent \textbf{Type of manuscript:} Original research article.

\vspace{0.1in}

\vspace{0.1in}
\noindent \textbf{Running head:} Generalizability analyses with a partially nested trial design

\vspace{0.1in}
\noindent \textbf{Conflicts of interest/Competing interests:} The authors have no relevant financial or non-financial interests to disclose.

\vspace{0.1in}
\noindent \textbf{Funding:} Removed for peer review. 

\vspace{0.1in}
\noindent \textbf{Author contribution:} All authors were involved in drafting the manuscript and have read and approved the final version submitted. No statistical analyses are included in the paper.

\vspace{0.1in}
\noindent \textbf{Data and computing code availability:} Not applicable. 

\vspace{0.1in}
\noindent \textbf{Consent to Participate/Publish (Ethics):} This research did not involve human subjects.

\vspace{0.1in}
\noindent
\textbf{Word count:} abstract = 196; main text $\approx$ 3900.

\vspace{0.1in}
\noindent
\textbf{Abbreviations:} No abbreviations used.

\vspace{0.1in}
\noindent
\textbf{Funding:} This work was supported in part by Patient-Centered Outcomes Research Institute (PCORI) awards ME-1502-27794, ME-2019C3-17875, and ME-2021C2-22365, and National Library of Medicine (NLM) award R01LM013616. The content of this paper is solely the responsibility of the authors and does not necessarily represent the official views of PCORI, the PCORI Board of Governors, the PCORI Methodology Committee, NLM, or the CASS investigators.

\thispagestyle{empty}

\clearpage
\thispagestyle{empty}

\vspace*{0.2in}

\begin{abstract}
\noindent
\linespread{1.5}\selectfont
We discuss generalizability analyses under a partially nested trial design, where part of the trial is nested within a cohort of trial-eligible individuals, while the rest of the trial is not nested. This design arises, for example, when only some centers participating in a trial are able to collect data on non-randomized individuals, or when data on non-randomized individuals cannot be collected for the full duration of the trial. Our work is motivated by the Necrotizing Enterocolitis Surgery Trial (NEST) that compared initial laparotomy versus peritoneal drain for infants with necrotizing enterocolitis or spontaneous intestinal perforation. During the first phase of the study, data were collected from randomized individuals as well as consenting non-randomized individuals; during the second phase of the study, however, data were only collected from randomized individuals, resulting in a partially nested trial design. We propose methods for generalizability analyses with partially nested trial designs. We describe identification conditions and propose estimators for causal estimands in the target population of all trial-eligible individuals, both randomized and non-randomized, in the part of the data where the trial is nested, while using trial information spanning both parts. We evaluate the estimators in a simulation study.
\end{abstract}

\clearpage
\section*{INTRODUCTION}
\setcounter{page}{1}

We discuss generalizability analyses under a partially nested trial design, where part of the trial is nested within a cohort of trial-eligible individuals, while the rest of the trial is not nested. This design arises, for example, when only some centers participating in the trial are able to collect data on non-randomized trial-eligible individuals or when data on non-randomized individuals cannot be collected for the full duration of the trial. We show that the partially nested trial design requires different identifiability assumptions and poses different modeling challenges compared with the more extensively studied \cite{cole2010, westreich2017, rudolph2017,tipton2018review, dahabreh2018generalizing, dahabreh2020transportingStatMed} fully nested or non-nested designs \cite{dahabreh2021studydesigns}.

Our work is motivated by the Necrotizing Enterocolitis Surgery Trial (NEST) \cite{blakely2021initial} that compared initial laparotomy versus peritoneal drain for infants with necrotizing enterocolitis or spontaneous intestinal perforation. In NEST, eligible infants who did not enter the randomized trial could be enrolled into an observational study. The NEST investigators originally planned to conduct a comprehensive cohort study \cite{olschewski1985,olschewski1992, schmoor1996} that would have resulted in a (``fully'') nested trial design \cite{dahabreh2021studydesigns}, where the entire trial would be prospectively embedded within a cohort of trial-eligible infants \cite{blakely2022special, shastry2022we, rich2012enrollment}. The conduct of the study followed this plan up to year 3 (from 2010 to 2013) but deviated from it when the investigators decided to stop enrollment into the observational component of the study, while continuing enrollment into the trial for the reminder of the study period (from 2014 to 2017). The investigators ended enrollment into the observational component due to concerns that providing the option to enroll into the observational component of the study was reducing the randomized trial enrollment and due to negative budgetary impacts from higher enrollment rates than expected into the observational component \cite{blakely2022special}. This resulted in a partially nested trial design because only the first phase of NEST was embedded within a cohort of trial-eligible individuals, but the second phase was not.
 
Here, using NEST as a motivating example, we propose novel methods for generalizability analyses that can be employed under partially nested trial designs. We obtain identification results for causal estimands in the target population of all trial-eligible individuals, both randomized and non-randomized, in the part of the data where the trial is nested, while using information spanning both parts of the trial. We show that in the partially nested trial design the lack of data on non-randomized individuals from the part of the study where there is no nesting complicates the modeling of the probability of trial participation and the probability of being in the part that has nesting, but does not preclude the identification of the causal estimands of interest. We propose estimators for these causal estimands and evaluate the finite-sample performance of the estimators in a simulation study.

\section*{STUDY DESIGN, DATA, ESTIMANDS} 

\paragraph{Study design and data:} Let $X$ be a vector of baseline (pre-randomization and pre-treatment) covariates; $S$ an indicator for trial participation (1 for trial participants; 0 for participants in the observational component); $A$ the treatment (randomly assigned in the trial); and $Y$ an outcome measured at the end of follow-up (continuous, binary, or count). In NEST \cite{blakely2022special}, $A$ is initial laparotomy or peritoneal drain and $Y$ is the primary composite outcome of death or neurodevelopmental impairment at 18 to 22 months corrected age (post term) according to standard practices within the Neonatral Research Network \cite{newman2012improving}.

\begin{figure}[!ht]
    \centering
    \caption{Data structures}
    \label{fig:study_design}
    \includegraphics[width=17cm]{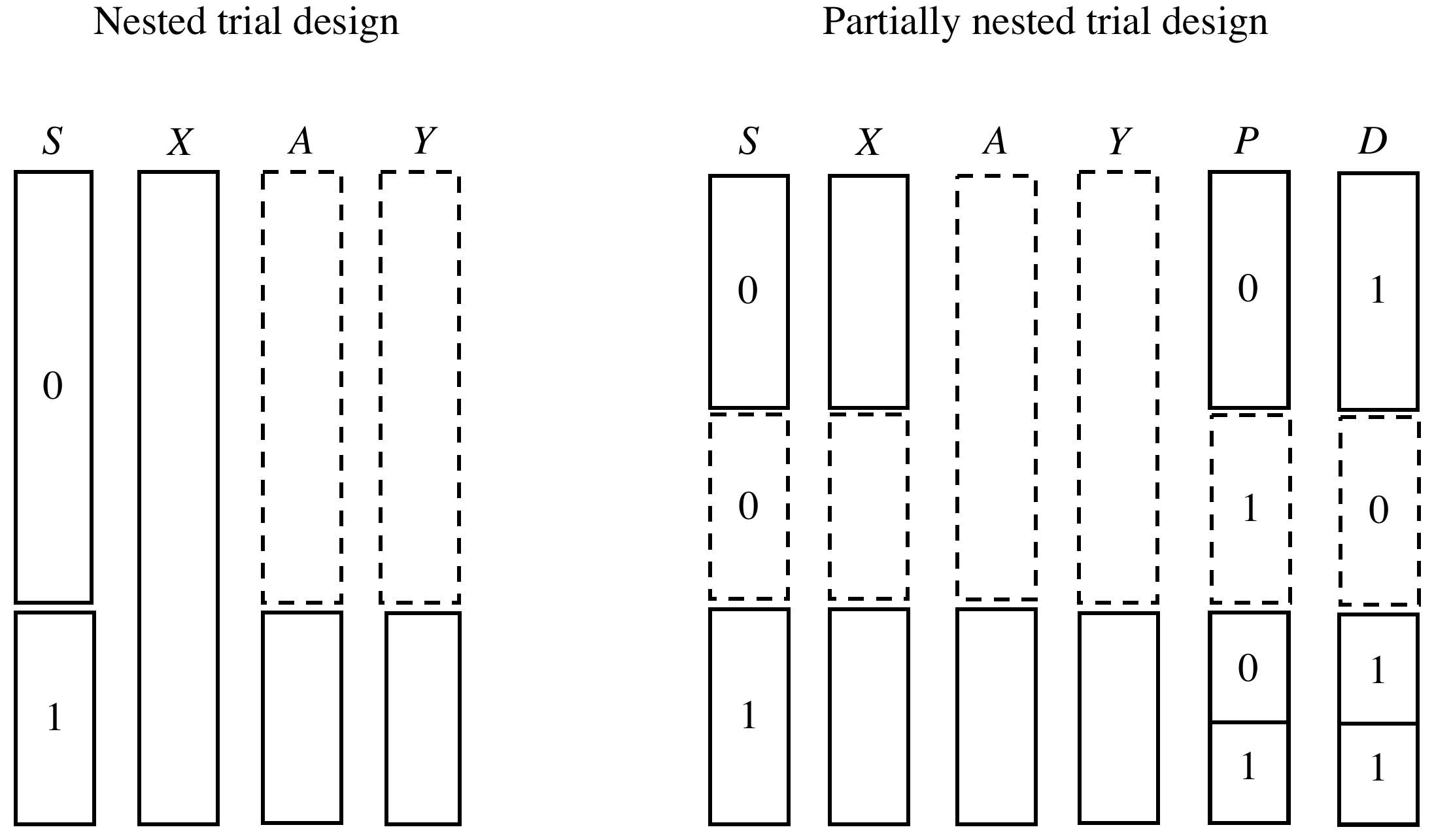}
    \caption*{The data structure of a nested trial design is shown on the left and the data structure of a partially nested trial design is shown on the right. \\
    For the data structure of both designs, $S=1$ indicates participation in the trial and $S=0$ indicates participation in the observational study. The $S=0$ bar is larger to indicate that in most datasets, the observational study is larger than the trial. $X$ indicates baseline covariates; $A$ indicates treatment, $Y$ indicates the outcome ($A$ and $Y$ need not be available among the $S=0$, indicated by the dashed lines). \\
    In the partially nested trial design, $P=0$ indicates the part of the data where the trial is nested; $P=1$ indicates the part of the data where the trial is not nested. $D=1$ indicates observations were sampled (randomized or non-randomized, regardless of the part of the study where they belong) and $D=0$ indicates observations not sampled into the data. \\
   The dashed lines in $P=1$ indicate missing data for $S=0$ in the partially nested trial design.}
\end{figure}

Figure \ref{fig:study_design} shows a schematic of the partially nested trial design and compares it with the previously proposed (fully) nested trial design \cite{dahabreh2021studydesigns}. Let $P$ be an indicator for the part of the data where there is nesting of trial participants in a cohort of trial-eligible individuals; $P=0$ indicates the part of the data where the trial is nested; $P=1$ indicates the part of the data where the trial is not nested. In NEST, $P$ is an indicator for the phase of enrollment: $P=0$ indicates the first phase, when information was collected from both randomized and non-randomized individuals; $P=1$ indicates the second phase, when information was collected only from randomized individuals. Because enrollment in the observational component ended early, there are no non-randomized individuals in the second phase (i.e., no individuals with $P=1$ and $S=0$). 

We model the part of the data where the trial is nested ($P=0$) as independent draws of the random tuple $O_{0,i} = (X_i, P_i=0, S_i, S_i A_i, S_i Y_i)$, $i = 1, \ldots, n_0$, where $n_0$ is the number of individuals in the part with $P=0$ (e.g., the first phase of NEST). We model the part of the data where the trial is not nested ($P=1$) as independent draws of the random tuple $O_{1,i} = (X_i, P_i=1, S_i=1, A_i, Y_i)$, $i = 1, \ldots, n_1$, where $n_1$ is the number of individuals in the part with $P=1$ (e.g., the second phase of NEST). We define the total sample size as $n = n_0 + n_1$.

Even though information on treatments and outcomes was available from non-randomized individuals in NEST, the methods we propose in the following sections require only baseline data from non-randomized individuals. We may not want to use outcome and treatment information from non-randomized individuals, even when available, for example, if the treatment - outcome association is intractably confounded outside the trial. Furthermore, to focus on issues of generalizability, we assume there is complete follow-up and adherence to the assigned treatment \cite{draftadherence}. In NEST, 295 of 308 (95.8\%) randomized infants had complete follow-up for the primary composite outcome of death or neurodevelopmental impairment, and 301 of 308 (97.7\%) infants adhered to the assigned treatment \cite{blakely2021initial}. The methods we propose can be extended to address these complications using well-understood approaches for trials with loss-to-followup or non-adherence \cite{robins2009, dahabreh2019identification}.

\paragraph{Sampling properties:} To understand the sampling scheme underlying partially nested trial designs, it helps to think of this design as obtained by selection from a fully nested trial design. Let $D$ be an indicator variable that takes the value 1 for sampled observations (randomized or non-randomized, regardless of the part of the study where they belong), or 0 for observations not sampled into the data. In the part of the data where the trial is nested ($P=0$), we have $\Pr[D=1|P=0]=1$ because all individuals, both randomized and non-randomized, are sampled. In the part of the data that does not have nesting ($P=1$), we have $\Pr[D=1|P=1, S=1] = 1$ because all randomized individuals are sampled, but $\Pr[D=1|P=1, S=0] = 0$ because no non-randomized individuals are sampled from the part of the study that does not have nesting. Thus, $\Pr[D=1|P=1]$ is equal to an unknown constant (but less than 1), because we do not know the number of non-randomized individuals in the part of the study that does not have nesting (i.e., we do not assume that information from individuals with $P=1, S=0$ is available).

\paragraph{Estimands:} The investigators of NEST  \cite{placeholder_rysavy2022draft} determined that the target population for generalizability analyses would comprise all infants, randomized and non-randomized, in the first phase of NEST. In other words, they viewed all infants with $P = 0$ as  representative of a clinically relevant population of trial-eligible infants who contributed information to the study. For that reason, we primarily focus on causal estimands that pertain to the target population that underlies the part of the data where the trial is nested ($P=0$). 

To define these causal estimands, we will use $Y^a$ to denote the counterfactual (potential) outcome under intervention to set treatment $A$ to $a$. The main estimands of interest will be expectations of the counterfactual outcome under intervention to set treatment $A$ to $a$ in the target population underlying the part of the data where the trial is nested ($P=0$), that is, $\E[Y^a | P = 0]$, for each $a$ in the (finite) set of treatments under consideration. These expectations are components of treatment effect measures; for example, the average treatment effect comparing intervention to set treatment to $a=1$ versus $a=0$ in the target population, is equal to the difference of the expectations of the counterfactual outcome under interventions $a=1$ and $a=0$: $\E[Y^{a=1} - Y^{a=0} | P = 0] = \E[Y^{a=1} | P = 0] - \E[ Y^{a=0} | P = 0]$.

\section*{IDENTIFICATION ANALYSIS}

\paragraph{Identifiability conditions:} We will argue that the following conditions are sufficient to identify the expectations of the counterfactual outcome under intervention to set treatment $A$ to $a$ in the target population, $\E[Y^a | P = 0]$:

\noindent
\emph{A1. Consistency of potential outcomes:} For every individual, $i$, whether randomized or non-randomized, and each $a \in \mathcal A$, if $A_i = a$, then $Y_i = Y^{a}_i$.

\noindent
\emph{A2. Mean exchangeability in the trial over $A$:} For every $a \in \mathcal A$ and every $x$ with positive density in the trial $f(x, S = 1 ) \neq 0$, $\E [ Y^{a} | X = x , S = 1, A =a] = \E [ Y^{a} | X = x, S = 1]$. 

\noindent
\emph{A3. Positivity of treatment assignment in the trial:} $\Pr[A=a | X = x, S=1] > 0$, for each $a \in \mathcal A$ and each $x$ with positive density in the trial $f(x, S = 1) \neq 0$.

\noindent
\emph{A4. Mean exchangeability between the trial and population underlying the part of the data where the trial is nested:} $\E [ Y^a | X = x, S = 1] = \E [ Y^a | X = x, P = 0 ] $, for every $x$ with positive density $f(x, P=0) \neq 0$ and for each $a \in \mathcal A$.

\noindent
\emph{A5. Positivity of trial participation:} $\Pr[S=1 | X = x] >0,$ for every $x$ such that $f(x, P=0) \neq 0.$

\paragraph{Reasoning about the identifiability conditions:} Condition A1 assumes the intervention is well-defined over all individuals in the target population and that there are no direct effects of trial participation on the outcome (i.e., no Hawthorne effects) \cite{dahabreh2018generalizing, dahabreh2019identification}.

Conditions A2 and A3 are expected to hold by design in trials because randomization of the treatment assignment ensures mean exchangeability (i.e., no unmeasured confounding) and positivity (i.e., every randomized individual has a non-zero probability of receiving treatment). Note that condition A2 follows from the fact that the randomized component of NEST used marginal randomization with a randomization ratio that did not change over the phases of the study, such that $(Y^a, X, P) \indep A | S = 1$. This independence condition implies that $Y^a \indep A | (X, S = 1)$, which in turn implies condition A2. 

Condition A4 states that the randomized participants are exchangeable (in expectation) with the target population underlying the part of the data where the trial is nested, conditional on baseline covariates. In Appendix A we discuss two ``more basic'' conditions that imply condition A4: conditional exchangeability between the two parts of the trial and conditional exchangeability between trial participants and non-participants in the part of the study where the trial is nested, and show that the former of these conditions has testable implications. Note in passing that condition A4 can be weakened if we are only interested in the average treatment effect, but not its component expectations, under a condition of exchangeability in effect measure over trial participation (e.g., see \cite{dahabreh2018generalizing} for analogous results in the fully nested trial design). 

Condition A5 states that, in large samples, every covariate pattern that can occur in the part of the data where the trial is nested will occur in the trial. This condition is in principle testable because it only involves the observed variables; when $X$ is high-dimensional, however, it is difficult to assess \cite{petersen2012diagnosing}.

\paragraph{Identification:} In the Appendix, we show that under assumptions A1 through A5, the expectation of the counterfactual outcome under intervention to set treatment $A$ to $a$ in the target population, $\E[Y^a | P = 0]$, is identified by the observed data functional 
\begin{equation}\label{eq:identification_g_form}
    \begin{split}
        \psi(a) &= \E \big[ \E[Y | X, S = 1, A =a ] \big | P = 0  \big] \\
        &=  \int \E[Y | X = x, S = 1, A =a ] f(x | P = 0) dx
    \end{split}
\end{equation}
This identification result suggests that the expectation of the observed outcome given covariates $X$ among trial participants who received treatment $a$ (the inner expectation) can be ``averaged'' (marginalized) over the covariate distribution of the target population (the outer expectation is conditional on $P = 0$). The result can be directly used in the partially nested trial design because under this sampling design (1) all trial participants are sampled, such that $\E[Y | X, S = 1, A =a ] = \E[Y | X, S = 1, A = a, D = 1] $; and (2) all individuals with $P=0$ are sampled, such that $f(X| P = 0) = f(X| P = 0, D= 1)$. 

In the Appendix, we also show that $\psi(a)$ has an algebraically equivalent weighting re-expression: 
\begin{equation}\label{eq:identification_weighting}
    \psi(a) = \dfrac{1}{\Pr[ P = 0 ]} \E \left[  \dfrac{I(S = 1, A = a) Y \Pr[P = 0 | X] }{ \Pr[S = 1 | X] \Pr[A = a | X, S = 1] }  \right].
\end{equation}
Leaving the proportionality constant $1/\Pr[P = 0]$ aside, it is useful to examine the weight component $\dfrac{\Pr[P = 0 | X]}{\Pr[S = 1 | X] \Pr[A = a | X, S = 1]}$ in the expectation of the above expression. The denominator is what we would expect from prior work on nested trial designs (e.g., \cite{dahabreh2018generalizing}) and serves to ``adjust'' for differences between the randomized trial and the target population, via the $\Pr[S = 1 | X]$ term, and for differences due to sampling variability between the treatment groups in the trial, via the $\Pr[A = a | X, S = 1]$ term. The numerator, $\Pr[P = 0 | X]$, ``calibrates'' the estimated treatment effect to the target population with $P=0$.

The identification result in equation \eqref{eq:identification_weighting}, however, cannot be directly used because $\Pr[P = 0]$, $\Pr[S = 1 | X],$ $\Pr[P = 0 | X]$, and the expectation term of the product are not identifiable under the partially nested design. These quantities are not identifiable because the number of non-randomized individuals with $P=1$ is not known and data from such individuals are not collected. In the data, we can only identify analogs of the terms in equation \eqref{eq:identification_weighting} conditional on $D=1$ (e.g., we can identify $\Pr[P = 0|D=1]$, which is not equal to $\Pr[P = 0 ]$).

In the Appendix, under the sampling properties of the partially nested trial design, we show that the ratio of the conditional probabilities in the observed data is identified \begin{equation*}
    \begin{split}
       \dfrac{\Pr[P=0|X,D=1]}{\Pr[S=1|X,D=1]} &=\dfrac{\Pr[P=0|X]}{\Pr[S=1|X]},
    \end{split}
\end{equation*} 
even though the numerators and denominators of the two fractions are not equal, $\Pr[P=0|X,D=1] \neq \Pr[P=0|X]$ and $\Pr[S=1|X,D=1] \neq \Pr[S=1|X]$. Furthermore, using the result above, properties of expectations, and properties of the sampling design, we also show that $$\psi(a) = \dfrac{1}{\Pr[ P = 0 | D = 1 ]} \E \left[  \dfrac{I(S = 1, A = a) Y \Pr[P = 0 | X, D = 1] }{ \Pr[S = 1 | X, D  =1 ] \Pr[A = a | X, S = 1] } \Big | D = 1  \right].$$ This result suggests that we can use weighting approaches in the observed data under the partially nested trial design (conditional on $D=1$), without needing information from non-randomized individuals in the part of the data that does not have nesting ($S=0,P=1$ and $D =0$).

\paragraph{Estimands of interest for another definition of the target population:} In some cases, we may be interested in the target population of all trial-eligible individuals, both randomized and non-randomized, in the population underlying both parts of the study (e.g., when data on non-randomized trial-eligible individuals could be at least \emph{in principle} collected). In such cases, instead of the expectation of the counterfactual outcome under intervention to set treatment $A$ to $a$ in the the target population with $P = 0$, we may be interested in the expectation of the counterfactual outcome in the population of trial-eligible individuals underlying both parts of the study, $\E[Y^a].$ In NEST, this may be viewed as a reasonable target population because, even though individuals with $S=0, P=1$ were not recruited in the second phase of the study, we nonetheless can easily conceive that happening. Informally, we can view the observations with $S=0, P=1$ as missing data. In other cases, however, it may not be possible to conceive the population of trial-eligible individuals for both parts of the study and $\E[Y^a]$ may not be well-defined.

When the population underlying both parts of the study is well-defined, we consider the following two identification conditions: 

\noindent
\emph{A4$^\dagger.$ Mean exchangeability between the trial and the entire population underlying both parts of the study:} $\E[ Y^a | X = x, S=1 ] = \E[ Y^a | X = x ]$, for every $x$ with positive density $f(x, P = 0) \neq 0$ and for each $a \in \mathcal A$.

\noindent
\emph{A6. The covariate distribution in both parts of the study is stable: } $f(X| P = 0) = f(X| P = 1) = f(X)$.

In the Appendix, we show that Condition A4$^\dagger$ can be obtained by combining condition A4 and an additional condition that the conditional expectation of the counterfactual outcomes among non-randomized individuals in the part of the data where the trial is nested is exchangeable with non-randomized individuals in the part of the data that does not have nesting, $\E[Y^a | X = x, P = 0, S=0] = \E[Y^a | X = x, P=1, S=0]$. 

Condition A6 requires the covariate distribution to be stable over both parts of the study and is not testable using the observed data because covariate information is unavailable for individuals with $P=1$ and $S=0$. In NEST, we might worry about condition A6 if external evidence suggested that the case-mix of enrolled infants had changed significantly over the study period. 

In the Appendix, we show that under conditions A1 through A3, A4$^\dagger$, A5 and A6, $\E[Y^a]$ is identifiable by $\psi(a)$. In the Appendix, we also show that under an assumption that the marginal probability of participation in the trial remains stable over both parts of the study (the part where the trial is nested and the part where it is not), we can identify the density of the covariates among non-randomized individuals in the part of the study where the trial is \emph{not} nested.

\section*{ESTIMATION AND INFERENCE}

\paragraph{Outcome modeling and standardization (g-formula):} The identification result in equation \eqref{eq:identification_g_form} suggests the following outcome model-based standardization (g-formula) estimator:
\begin{equation}\label{eq:estimation_g_form}
    \widehat \psi_{\text{\tiny g}}(a) = \left\{ \sum\limits_{i=1}^{n} I(P_i = 0) \right\}^{-1} \sum\limits_{i=1}^{n} I(P_i = 0) \widehat g_a(X_i),
\end{equation}
where $ \widehat g_a(X)$ is an estimator of the expectation of the outcome conditional on baseline covariates among trial participants, $\E[Y | X, S = 1, A =a, D=1 ] =  \E[Y | X, S = 1, A =a]$. We model the expectation of the outcome separately in each treatment arm to more flexibly reflect heterogeneity. When the model is correctly specified such that $ \widehat g_a(X)$ is consistent for $\E[Y | X, S = 1, A =a]$,  then $\widehat \psi_{\text{\tiny g}}(a)$ converges in probability to $\psi(a).$

\paragraph{Weighting:}The identification result in equation \eqref{eq:identification_weighting} suggests the following weighting estimator: 
\begin{equation}\label{eq:estimation_weighting}
    \widehat \psi_{\text{\tiny w}}(a) = \left\{ \sum\limits_{i = 1}^{n} I(P_i = 0) \right\}^{-1} \sum\limits_{i = 1}^{n}  \widehat w_a(X_i) Y_i,
\end{equation}
where the weights $ \widehat w_a(X) $ are defined as 
\begin{equation*}
   \widehat w_a(X) = \dfrac{I(S = 1, A = a) \widehat q(X)}{\widehat p(X) \widehat e_a(X)},
\end{equation*}
where $\widehat q(X)$ is an estimator for the probability of being in the part of the data that where the trial is nested, $\Pr[P = 0 | X, D=1]$; $\widehat p(X)$ is an estimator for the probability of trial participation, $\Pr[S = 1 | X, D=1]$; and $\widehat e_a(X)$ is an estimator for the probability of treatment in the trial, $\Pr[A = a | X, S = 1,D=1] = \Pr[A = a | X, S = 1]$. Estimating the weights requires specifying models for these probabilities. The probability of treatment in the trial is known by design, so the model for $\Pr[A = a | X, S = 1]$  can always be correctly specified, or the known-by-design probability can be used instead. Nevertheless, estimating that probability may ``adjust'' for imbalances of baseline covariates in the trial and improve efficiency \cite{lunceford2004, williamson2014variance}.
When the necessary models are correctly specified, such that $\widehat q(X)$ is consistent for $\Pr[P = 0 | X, D=1]$ and $\widehat p(X)$ is consistent for $\Pr[S = 1 | X, D=1]$, then $\widehat \psi_{\text{\tiny w}}(a)$ converges in probability to $\psi(a)$.

\paragraph{Augmented weighting estimator:} Last, we propose an estimator that combines both outcome modeling and weighting: 
\begin{equation}\label{eq:estimation_dr}
    \widehat \psi_{\text{\tiny aug}}(a) = \left\{ \sum\limits_{i = 1}^{n} I(P_i = 0) \right\}^{-1} \sum\limits_{i = 1}^{n}  \Big\{ \widehat w_a(X_i) \big\{ Y_i - \widehat g_a(X_i) \big\} + I(P_i = 0) \widehat g_a(X_i) \Big\}.
\end{equation}
We refer to $\widehat \psi_{\text{\tiny aug}}(a)$ as an ``augmented weighting'' estimator because it can be viewed as a version of the weighting estimator that is augmented with an outcome model. In the Appendix, we show this estimator is model doubly robust: it is consistent when either the models for $\Pr[P = 0 | X, D=1]$, $\Pr[S = 1 | X, D=1]$, and $\Pr[A=a|X, S=1]$ are correctly specified, or the model for $\E[Y|X, S=1, A=a]$ is correctly specified (but not necessarily all four models) \cite{bang2005, smucler2019unifying}. Furthermore, if data-adaptive approaches (machine learning) are used to estimate the models, $\widehat \psi_{\text{\tiny aug}}(a)$ can allow for valid inference when the data-adaptive approaches have rates of convergence slower than the parametric rate\cite{chernozhukov2018double,smucler2019unifying}.

\paragraph{Inference:} We can obtain standard errors for the estimators in equations \eqref{eq:estimation_g_form}, \eqref{eq:estimation_weighting}, and \eqref{eq:estimation_dr} using M-estimation methods (i.e., the sandwich estimator) \cite{stefanski2002}. Alternatively, we can use bootstrap methods \cite{efron1994introduction}, which may be more convenient to implement. Both of these methods can properly account for uncertainty when estimating the models needed for different estimators.


\paragraph{Modeling for estimating the weights:} Modeling the probability ratio $\Pr[P = 0 | X , D = 1]/\Pr[S = 1 | X, D = 1]$, which is a key part of the weights, requires some care. One approach is to separately model the conditional probabilities in the numerator and denominator of the ratio. For example, we can use logistic regression models to estimate each term, richly parameterized to reduce model misspecification. In the Appendix we show in a simulated example that separately modeling the conditional probabilities in the numerator and denominator using flexible model specification can lead to reasonable approximations of the ratio of the two probabilities.

A useful model assessment can be based on the mean of the estimated weights, as was proposed in \cite{dahabreh2020transportingStatMed} for other study designs. Specifically, the identity $$ \E \left[  \dfrac{I(S = 1) \Pr[P = 0 | X,D=1] }{ \Pr[S = 1 | X,D=1] } \Big | D = 1  \right] = \Pr[P=0|D=1]$$ suggests the following diagnostic: $$ \sum\limits_{i = 1}^{n}    \dfrac{I(S_i = 1)\widehat q(X_i)}{\widehat p(X_i)}  \approx  \sum\limits_{i = 1}^{n} I(P_i = 0).$$ Large differences between the left and right hand sides of the expression above (when the ratio of the two terms is far from 1) suggest near-violations of condition A5 or misspecification of the models used to calculate the weights.

\section*{SIMULATION STUDY}

To examine the finite-sample performance of the estimator described above, we conducted a simulation study roughly based on the sample sizes and proportion of individuals selected into the trial or observational component of NEST. In NEST, 308 infants were enrolled in the randomized component of the study during both phases (156 in the first phase; 152 in the second phase) and 226 infants were enrolled in the observational component during the first phase. 

\paragraph{Baseline data generation:} We first simulated a fully nested trial design with 750 trial-eligible individuals. For each individual, we simulated  covariates, $X=(1, X^{(1)}, X^{(2)}, X^{(3)}), $ where $X^{(j)}$, $j=1,2,3$, had independent standard normal distributions. We then generated the binary indicator for trial participation, $S$, from a Bernoulli distribution with parameters $\Pr[S = 1 | X]= \dfrac{\text{exp}(\beta X^T) }{1+ \text{exp}(\beta X^T)}$, where $\beta = (-0.471, 0.5, 0.5,0.5 ).$ Using numerical methods \cite{robertson2022using}, we selected -0.471 as the value for the intercept, such that approximately 40\% of all simulated individuals were trial participants.

\paragraph{Inducing a partially nested trial design:} We generated the binary indicator for the part of the data that has nesting, $P$, from a Bernoulli distribution with parameter $\Pr[P = 1]=0.5$, such that 50 percent of the individuals are randomly enrolled in each part. Then, to form the partially nested trial design, we dropped the non-randomized individuals in the part of the study where there was no nesting of the trial (i.e., we dropped observations with $S = 0, P = 1$).

\paragraph{Treatment generation:} 
Among randomized individuals, we generated an indicator for treatment assignment $A$ from a Bernoulli distribution with parameter $\Pr [A = 1] = 0.5 $.

\paragraph{Outcome generation:}  We generated counterfactual outcomes for $a=1,0$ as $Y^a$ from treatment-specific Bernoulli distributions with parameters $\Pr[Y^a = 1 | X]= \dfrac{\text{exp}(\zeta^a X^T) }{1+ \text{exp}(\zeta^a X^T)}$, for $a=0,1$. We varied the coefficients of $\zeta^1$ to evaluate scenarios with strong, moderate, and no effect modification on the log-odds ratio scale. For all scenarios we used  $\zeta^0=(0.5,0.5,0.5,0.5).$ For strong effect modification, $\zeta^1=(1,0,0,0.5)$; moderate effect modification, $\zeta^1=(1,0,0.5,0.5)$; and no effect modification, $\zeta^1=(1,0.5,0.5,0.5)$. We generated the observed outcome for randomized individuals using the consistency assumption: $Y = A Y^1 + (1-A) Y^0.$ 

\paragraph{Estimators:} We implemented the proposed estimators -- the g-formula estimator in equation \eqref{eq:estimation_g_form}, the weighting estimator in equation \eqref{eq:estimation_weighting}, and the augmented weighting estimator in equation \eqref{eq:estimation_dr} -- using models in the simulated data to estimate the expectations of the counterfactual outcome under different treatments (i.e., treatment-specific risks) and the average treatment effect (i.e., the risk difference) in the population underlying the part of the data where the trial was nested ($P=0$). Specifically, we used logistic regression models for the outcome, the probability of participation in the trial, the probability of being in the part of the data where the trial is nested, and the probability of treatment in the trial, conditional on the linear main effects of the baseline covariates. For comparison, we implemented a trial-only estimator that used the same outcome model, but averaged the model predictions only over the sample of trial participants (this g-formula estimator is a covariate-adjusted estimator of the treatment effect in the population underlying the trial, based on a correctly specified model). In our simulation, estimates from the trial-only estimator will be biased for the average treatment effect in the target population underlying the part of the data where the trial was nested because there is effect modification (on the logit) scale and selective participation in the trial (resulting in a different distribution of effect modifiers in the trial and the target population).

\paragraph{Performance measures:} We evaluated the performance of the proposed estimators by calculating their mean bias, variance, and coverage over 1000 runs. We obtained the true value in the target population using Monte Carlo methods. We evaluated the coverage of nominal 95\% confidence intervals obtained using standard errors from M-estimation and the bootstrap (1000 bootstrap runs in each simulation run). To implement M-estimation, we used the \texttt{geex} package \cite{saul2020calculus} in \texttt{R} \cite{currentRcitation}. 

\paragraph{Results:} As expected, the trial-only estimator was biased for estimating the expectations of the counterfactual outcome and the average treatment effect in the target population. All other proposed estimators were approximately unbiased (see Table \ref{Table:sim_biasmain}). Table \ref{Table:sim_sdmain} shows that these estimators had similar estimated standard deviation and Table \ref{Table:sim_coveragemain} shows that they had similar coverage with either the sandwich estimator or bootstrap estimator of the variance.  

\paragraph{Additional simulation scenarios and results:} We repeated the simulation but replaced the binary outcome with a continuous outcome, such that $Y^a=\zeta^a X^T + \epsilon^a,$ where we generated the errors $\epsilon^a,$ for $a=1,0$ from independent standard normal distributions. Full results are reported in the Appendix. We found that the performance was similar to the simulation in the main text with the binary outcome.

\section*{DISCUSSION}

Motivated by NEST \cite{blakely2021initial}, we developed methods for partially nested trial designs that allow learning about treatment effects in the population underlying the part of the data where the trial is nested, using treatment and outcome information from the entire trial. We provided identification results that do not require information from non-randomized individuals in the part of the data that does not have nesting. These results rely on different causal and sampling assumptions compared with fully nested or non-nested trial designs \cite{cole2010, westreich2017, rudolph2017,tipton2018review, dahabreh2018generalizing, dahabreh2020transportingStatMed}. We also described challenges when modeling the ratio of the probability of trial participation to the probability of being in the part that has has nesting; these challenges are unique to partially nested trial designs. We also proposed different estimators for this design and found that they had good finite-sample performance in a simulation study. 

Partially nested trial designs may arise in a variety of practical settings. In NEST, the design arose because the investigators chose to end enrollment into the observational component of a comprehensive cohort study early, such that only the first phase of the study exhibited nesting of the trial within the cohort of trial-eligible individuals. In multi-center trials, the partially nested design can arise when some centers are unable or unwilling to collect data from non-randomized individuals. In studies using data linkage methods to retrospectively nest a trial in registry or routinely collected data, the partially nested design can arise when linkage is incomplete or when data access is restricted, such that only part of the trial can be successfully linked (and, thus, nested) in the registry or routinely collected data. 

In some cases, alternatives to partial nesting are possible. For example, when data from both trial-participants and non-participants are collected prospectively, investigators can subsample non-participants to improve research economy without sacrificing generalizability \cite{dahabreh2021studydesigns}. In NEST, for instance, instead of terminating enrollment into the observational component of the study early, it might have been possible to randomly subsample non-randomized infants over the entire study period. Such a design might have allowed the estimation of treatment effects in the entire target population (not just the population underlying the early phase of NEST) while requiring similar resources as the partially nested design.

In sum, we have proposed methods that can be used for generalizability analyses for partially nested designs. This novel design variant highlights the importance of jointly considering study design (and its attendant sampling properties), causal assumptions, and statistical models when combining information from trials and additional data sources to answer causal questions.

\clearpage
\renewcommand{\refname}{REFERENCES}
\bibliographystyle{ieeetr}
\bibliography{bibliography}

\clearpage
\section*{TABLES}

\begin{table}[ht]
\centering
 \caption{Scaled bias results.}
\label{Table:sim_biasmain}
\begin{tabular}{@{}lccccc@{}}
\toprule
Scenario    & Estimand & Trial  &  $\widehat \psi_{\text{\tiny g}}$     &  $\widehat \psi_{\text{\tiny w}}$    &  $\widehat \psi_{\text{\tiny aug}}$   \\ \midrule
No EM       & $\E[Y^0 | P = 0]$  & 2.212  & 0.017  & 0.034  & 0.014  \\
            & $\E[Y^1 | P = 0]$ &      1.911  & -0.005 & 0.016  & 0.000  \\
            & ATE    & -0.301 & -0.022 & -0.018 & -0.014 \\ \midrule
Moderate EM & $\E[Y^0 | P = 0]$       & 2.163  & -0.024 & 0.000      & -0.026 \\
            & $\E[Y^1 | P = 0]$      & 1.318  & 0.011  & 0.007  & 0.014  \\
            & ATE     & -0.844 & 0.035  & 0.007  & 0.039  \\ \midrule
Strong EM   & $\E[Y^0 | P = 0]$      & 2.177  & -0.021 & -0.006 & -0.023  \\
            & $\E[Y^1 | P = 0]$      & 0.723  & 0.033  & 0.056  & 0.050  \\
            & ATE     & -1.453 & 0.055  & 0.062  & 0.073 \\
 \bottomrule
\end{tabular}
\caption*{Results are scaled (multiplied by $\sqrt{750}$). \\
EM = effect modification on the logit scale; ATE = $\E[Y^1|P=0]-\E[Y^0|P=0];$ Trial = trial-only estimator; $\widehat \psi_{\text{\tiny g}}$  = outcome-modeling estimator in equation \eqref{eq:estimation_g_form}; $\widehat \psi_{\text{\tiny w}}$  = weighting estimator in equation \eqref{eq:estimation_weighting}; $\widehat \psi_{\text{\tiny aug}}$ = augmented weighting estimator in equation \eqref{eq:estimation_dr}.}
\end{table}

\begin{table}[ht!]
\centering
 \caption{Scaled estimated standard deviation results.}
\label{Table:sim_sdmain}
\begin{tabular}{@{}lcccc@{}}
\toprule
Scenario    & Estimand &   $\widehat \psi_{\text{\tiny g}}$     &  $\widehat \psi_{\text{\tiny w}}$    &  $\widehat \psi_{\text{\tiny aug}}$  \\ \midrule
No EM       & $\E[Y^0 | P = 0 ]$       & 1.161       & 1.215       & 1.216   \\
            & $\E[Y^1 | P = 0 ]$      & 1.190        & 1.228       & 1.237  \\
            & ATE      & 1.671       & 1.71        & 1.723  \\ \midrule
Moderate EM & $\E[Y^0 | P = 0 ]$        & 1.231       & 1.276       & 1.268  \\
            & $\E[Y^1 | P = 0 ]$       & 1.201       & 1.243       & 1.235  \\
            & ATE     & 1.631       & 1.701       & 1.689  \\ \midrule
Strong EM   & $\E[Y^0 | P = 0 ]$        & 1.174       & 1.209       & 1.206  \\
            & $\E[Y^1 | P = 0 ]$      & 1.114       & 1.211       & 1.153 \\
            & ATE       & 1.618       & 1.731       & 1.671  \\ \bottomrule
\end{tabular}
\caption*{Results are scaled (multiplied by $\sqrt{750}$). \\
EM = effect modification on the logit scale; ATE = $\E[Y^1|P=0]-\E[Y^0|P=0];$ $\widehat \psi_{\text{\tiny g}}$  = outcome-modeling estimator in equation \eqref{eq:estimation_g_form}; $\widehat \psi_{\text{\tiny w}}$  = weighting estimator in equation \eqref{eq:estimation_weighting}; $\widehat \psi_{\text{\tiny aug}}$ = augmented weighting estimator in equation \eqref{eq:estimation_dr}.}
\end{table}

\begin{table}[ht!]
\centering
 \caption{Coverage.}
\label{Table:sim_coveragemain}
\begin{tabular}{@{}lccccccc@{}}
\toprule
Scenario & Estimand & \multicolumn{3}{c}{Sandwich} & \multicolumn{3}{c}{Bootstrap} \\ \midrule
         &         &  $\widehat \psi_{\text{\tiny g}}$     &  $\widehat \psi_{\text{\tiny w}}$    &  $\widehat \psi_{\text{\tiny aug}}$      &  $\widehat \psi_{\text{\tiny g}}$     &  $\widehat \psi_{\text{\tiny w}}$    &  $\widehat \psi_{\text{\tiny aug}}$     \\
No EM       & $\E[Y^0 | P = 0 ]$       & 0.950           & 0.952          & 0.947           & 0.956          & 0.941          & 0.953      \\
        & $\E[Y^1 | P = 0 ]$        & 0.929          & 0.932          & 0.929           & 0.931          & 0.940           & 0.932    \\
        & ATE      & 0.945          & 0.944          & 0.937           & 0.946          & 0.948          & 0.945   \\  \midrule
Moderate EM        & $\E[Y^0 | P = 0 ]$        & 0.937          & 0.938          & 0.931           & 0.937          & 0.930           & 0.938    \\
        & $\E[Y^1 | P = 0 ]$       & 0.928          & 0.932          & 0.925           & 0.928          & 0.940           & 0.930     \\
       & ATE      & 0.945          & 0.945          & 0.941           & 0.949          & 0.956          & 0.950   \\  \midrule
Strong EM        & $\E[Y^0 | P = 0 ]$        & 0.941          & 0.947          & 0.944           & 0.948          & 0.957          & 0.950  \\
        & $\E[Y^1 | P = 0 ]$        & 0.940           & 0.953          & 0.931           & 0.942          & 0.957          & 0.940     \\
        & ATE      & 0.944          & 0.943          & 0.937           & 0.946          & 0.951          & 0.942     \\  
 \bottomrule
\end{tabular}
\caption*{
EM = effect modification on the logit scale; ATE = $\E[Y^1|P=0]-\E[Y^0|P=0];$ $\widehat \psi_{\text{\tiny g}}$  = outcome-modeling estimator in equation \eqref{eq:estimation_g_form}; $\widehat \psi_{\text{\tiny w}}$  = weighting estimator in equation \eqref{eq:estimation_weighting}; $\widehat \psi_{\text{\tiny aug}}$ = augmented weighting estimator in equation \eqref{eq:estimation_dr}.}
\end{table}

\clearpage
\section*{FIGURES}

\end{document}